\documentclass[conference]{IEEEtran}
\IEEEoverridecommandlockouts
\usepackage{cite}
\usepackage{amsmath,amssymb,amsfonts}
\usepackage{algorithmic}
\usepackage{graphicx}
\usepackage{textcomp}
\usepackage{xcolor}
\usepackage{calc}
\usepackage{cclicenses}
\usepackage{enumitem}
\setlist[description]{nosep, topsep = 1em, labelindent=0em, listparindent=4em, leftmargin=1em}%

\def\BibTeX{{\rm B\kern-.05em{\sc i\kern-.025em b}\kern-.08em
    T\kern-.1667em\lower.7ex\hbox{E}\kern-.125emX}}

\makeatletter
\def\ps@IEEEtitlepagestyle{%
	\def\@oddfoot{\mycopyrightnotice}%
	\def\@evenfoot{}%
}
\def\mycopyrightnotice{%
	{\footnotesize \copyright Alain Brenzikofer\hfill\hfill encointer.org\hfill first published on 18.11.2018, preliminary V0.13, 09.12.2020}% <--- Change here
	\gdef\mycopyrightnotice{}% just in case
}

\newcommand{\executeiffilenewer}[3]{%
	\ifnum\pdfstrcmp{\pdffilemoddate{#1}}%
	{\pdffilemoddate{#2}}>0%
	{\immediate\write18{#3}}\fi%
}

\newtheorem{erule}{Rule}
\newtheorem{hypothesis}{Hypothesis}
\begin{document}
%\SetWatermarkText{Confidential Draft}
%\SetWatermarkScale{0.5}
\newcommand{\encointer}{\textsl{Encointer} }

\title{\encointer - Local Community Cryptocurrencies with Universal Basic Income\\
}

\author{\IEEEauthorblockN{Alain Brenzikofer}
alain@encointer.org}
%\and

\maketitle

\begin{abstract}
\encointer proposes a blockchain platform for local community cryptocurrencies. Individuals can claim a universal basic income through issuance of fresh money. Money supply is kept in proportion to population size through the use of demurrage. Sybil attacks are prevented by regular, concurrent and randomized pseudonym key signing parties to obtain a proof-of-personhood. \encointer features privacy by design and purchasing-power adjusted transaction fees. 
\end{abstract}


\section{Motivation}

\subsection{Economics}

With the appearance of Bitcoin \cite{nakamoto08} in 2008, a big socio-economic experiment took off. The nature of money itself was widely debated. Bitcoin adopts a hard-coded nominally inflatonary monetary policy saturating at a fixed supply. Rapid adoption made Bitcoin a real deflationary currency, which it will remain if successful. Early adopters made a fortune. Because of its deflationary nature, bitcoin favors accumulation of capital for the few. Wealth increases without merit.

The monetary policy followed by central banks issuing national fiat money on the other hand often follows the goal of price stability, aiming at a moderate inflation goal in the order of 1-2\%. Issuance of money is appointed to banks who give credit to companies who employ workers who consume goods and thereby make companies profitable and raise the GDP. A process that allegedly benefits everyone. 
However, the observation that an increase in money supply doesn't benefit everyone equally is referred to as the Cantillon-Effect \cite{cantillon}. Thomas Piketty shows \cite{piketty} that gains on capital historically exceed economic growth, another factor that questions the trickle-down theory. 

\encointer aims at turning this logic upside-down and lets all individuals issue money subject to common rules. In order to mint \encointer, people need to attend pseudonym key signing parties (meetups) that happen at regular intervals at high sun all over the world within small randomized groups of people. The \encointer issuance therefore represents a form of \textit{universal basic income} (UBI) for every person attending such meetups. 

These \encointer meetups are at the same time the basis of a digital personhood claim, or proof-of-personhood (PoP) \cite{ford08} \cite{pop}, proving a one-to-one relationship between a person and her digital identity. One person can only maintain one personhood claim because ceremonies are designed to make it impossible to attend two meetups physically as they happen in different places concurrently. 

While all other cryptocurrencies to date are global units of value, \encointer enables local community currencies. Because its money issuance is bound to the physical presence of people at specified locations at a specified time, this issuance can be geographically bound to enable regional currencies with their independent valuation. Regional currencies have been proposed as more sustainable alternatives to national money \cite{gesell}\cite{lietaer}.

\subsection{Decentralization}

\encointer aims to be resilient against censorship, attacks and failures. Resilience shall be achieved through decentralization:

\begin{description}
	\item[architectural decentralization] avoid a single point of failure through redundancy (different machines in different geographic locations)
	\item[political decentralization] distribute control among many individuals and organizations
%	\item[logical decentralization] ensure that the system can continue to exist if it is divided into arbitrary subgroups
\end{description}

In order to achieve decentralization, Bitcoin and many other cryptocurrencies use an energy-hungry consensus mechanism called proof of work (PoW). While PoW has been the key idea that made Bitcoin possible in the first place, it is not ecologically sustainable. Moreover, it failed its goal of decentralization as mining has become centralized by a single company in a single country. 

Peercoin \cite{sunnyking12} introduced the first proof-of-stake (PoS) cryptocurrency in 2012. While PoW makes a compromise on energy efficiency, PoS makes the compromise of benefitting and empowering the rent-seeking wealthy.

PoPcoin \cite{pop} introduced a more democratic consensus algorithm where all persons with a proof-of-personhood hold an equal right to produce new blocks. PoP consensus, however, faces a chicken-egg situation: The consensus algorithm which should ensure the security of PoP relies on PoP itself. Once established, this consensus algorithm is the closest fit to our stated goals but it needs a stepping stone to be bootstrapped.

Polkadot \cite{polkadot} can serve as this stepping stone. It introduced a heterogeneous multi-chain framework with shared security, based on Nominated PoS (NPoS) consensus. \encointer aims to benefit from Polkadot's shared security by becoming a parachain until eventually migrating to PoP consensus.

\subsection{Transaction Privacy}
Bitcoin transactions are pseudonymous but not anonymous. It has been shown that identities of transacting parties can be revealed \cite{reid12}. Aiming at transaction privacy, Monero was introduced in 2014, employing the CryptoNote protocol \cite{saberhagen14}. Receiving funds in Monero means scanning every block for transactions to oneself. This task can only be taken out by full nodes as delegating it would leak private information. 

Zcash was introduced in 2016 employing the Zerocash protocol \cite{bensasson14} using zk-SNARKS to hide sender, receiver and value from third parties. Generating SNARKS to send funds is a computationally heavy process, limiting its usability for mobile and IoT devices.

For both Monero and ZCash, privacy comes at the price of large transaction size, letting the blockchain grow quickly.

Hyperledger Sawtooth has demonstrated Private Data Objects (PDO) \cite{sawtoothpdo}. PDOs allow to take state and execution of state transitions off chain. PDOs rely on trusted execution environments (TEE) and require trust in vendor attestation services. Currently, there are few TEE vendors on the market (i.e. Intel SGX \cite{costan16}, ARM trust zone \cite{trustzone} used by AMD, Qualcomm and others) but there are also open-source hardware initiatives that might one day diversify the attestation trust. 

SubstraTEE \cite{substraTEE} is a TEE-based framework for trusted off-chain computation and TEE-validated sidechains on a second layer offering confidentiality and scalability at the same time. \encointer builds on this framework to ensure private transactions.
 
\subsection{Scalability}

Because of its block size limit, Bitcoin can only reach about 4-7 transactions per second onchain. 
In order to tackle Bitcoin's scalability issues, Lightning Network payment channels \cite{poon15} were introduced in 2015 and demonstrated in 2017. Scalability is achieved by bilaterally treating transactions off-chain with the option to settle the last balance at any time on-chain. Teechan \cite{lind17} was introduced in 2017, implementing payment channels in TEEs. 

As mentioned in the previous section, \encointer takes the execution of state transitions off-chain altogether leveraging the SubstraTEE framework \cite{substraTEE}. With individual shards for each local community, \encointer can effectively scale horizontally while still maintaining interchangeability among local currencies.

\subsection{Governance}

Decentralized blockchain governance is the process of deciding on the future of the chain's rules, and any interventions necessary to avoid undesired effects. In the case of Bitcoin, a balance of power between miners and coin holders decides about the future of the protocol by putting their bet (mining power, bids on exchanges) on the desired option. This off-chain decision process lead to multiple chain forks in the past, hurting the ecosystem and dividing development teams. 

PoS blockchains delegate governance to their coin holders entirely. Polkadot \cite{polkadot} employs a very elaborate representative on-chain governance scheme with referenda. This process can avoid forks because all interventions are decided on-chain. While this design promotes a distribution of power, the underlying rule is still: Who has more coin has more say, a paradigm which is incompatible with our egalitarian ambition.

As \encointer has an anti-sybil attack measure in place (uPoP), a democratic one-person-one-vote (1p1v) scheme can be implemented within the boundaries of one local currency community. Deciding on the parameters of a local currency should be left to the community maintaining and using it.

Global protocol decisions can't simply be delegated to all persons from all local currency communities because the trust model doesn't reach beyond the currency that you yourself participate in (attackers can undetectably build bot communities as long as they never interact with honest humans). 

One way to extend the trust model are decentralized exchanges among \encointer currencies. Communities that buy each other's coins quantifiably trust the integrity of the other community. Even more trust comes from people traveling and are participating in different communities. If there is one community that acts as a root of trust, a global 1p1v scheme becomes possible.

\encointer appoints a root of limited trust to a Swiss association holding the \encointer trademark. 
The association defines and audits the root-of-trust community and suggests protocol updates in advance, including changes of rewards and fees or block size limits. Suggestions by the association can be blocked by a referendum vote requiring a majority of 2/3 of voters. Balloting happens on-chain anonymously. The 2/3 majority threshold for referendums allows the association to react quickly to changing circumstances but still provide decentralization, given large opposition.

If the \encointer association should fail to suggest necessary changes, the community may suggest changes as well. They also require a majority of 2/3 of stake and 2/3 majority of 1p1v voters.

Once a dense web-of-trust has been established, the association becomes obsolete in its function to act as a root of trust and it can be replaced by an on-chain elected council similar to Polkadot's - but in contrast to the latter it will be democratically legitimized.

\section{Local Currencies}
Encointer is no single currency. It manages an unpermissioned set of many local currencies. Every geographically bound community can have their own local currency.

\subsection{Bootstrapping a Local Currency}
\begin{figure}
	\centering
	\def\svgwidth{\columnwidth}
	\includegraphics[width=\columnwidth]{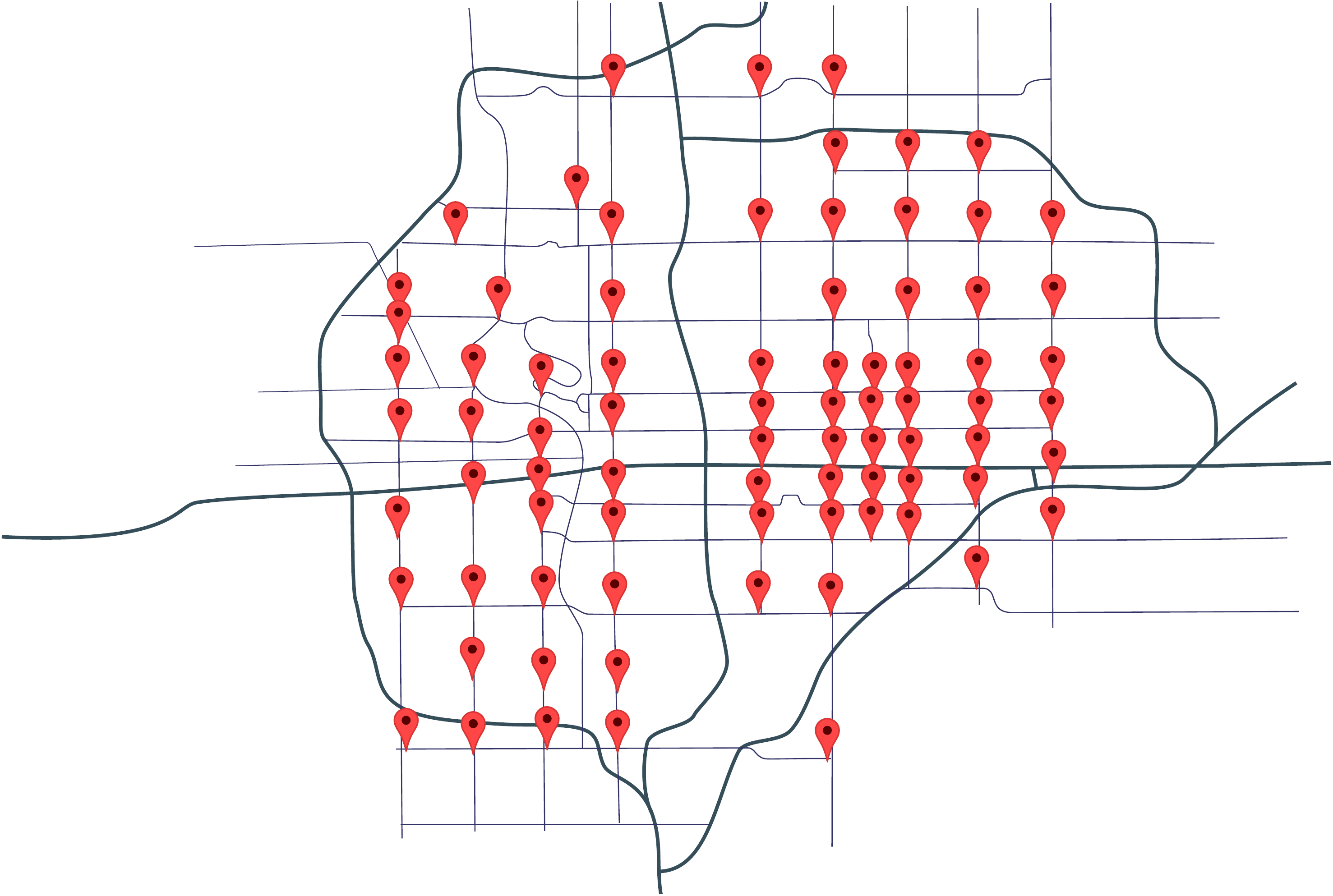}
	\caption{Possible public meetup locations $l_{C,q}$ for a local currency}
	\label{fig:map}
\end{figure}

Initiators of a new local \encointer currency need to define its geographical bound by a large set $L_C$ of possible publicly accessible meetup locations $l_{C,q}$ as shown in figure \ref{fig:map}. As people will have to attend some meetup at a random location from the set $L_C$, the geographical extension should be chosen reasonably such that people only need to travel within acceptable ranges. In order to improve meetup randomization, the set of possible meetup locations should be chosen larger than the region's population size. The location set can later be modified by on-chain governance only. 

Then the initiators need to perform a trusted setup ceremony with 3-12 participants. An \encointer local currency should only be trusted if it was bootstrapped with a public trusted setup ceremony. In the best case you find locally renowned people to participate in the first ceremony. For the trusted setup it is recommended that public keys used for the ceremony are made public by their owners.

The currency's identifyer $C_k$ is the group public key of all registered trusted setup participants.

\subsection{Urban Scalability}
The most densely inhabitated city currently is Manila with $43'000/km^2$ \cite{manila18}. With a limit of 10 people per meetup and a participation of 100\% , $4300$ meetups would take place per $km^2$. Each meetup would be allowed an area of $232 m^2$ and the distance between meetup locations would be $r_{s,i} \approx 15m$. 
%The time window $\delta_T{s,i}$ would be as low as $180ms$, which is achievable by automated witnessing over i.e. bluetooth.

\section{Unique PoP Ceremonies}\label{sec:upop}
Key signing ceremonies will be scheduled every 41 days. The interval of 41 days is an arbitrary design choice. In order for many people to be able to participate, it should be at different weekdays every time and it shouldn't happen too often - but often enough that missing one ceremony doesn't hurt too much. 
All meetups will happen at high sun $T_{C,i}$ at the same date all over the world. This is crucial because we want nobody to be able to attent two meetups for the same ceremony, because this would allow a single person to maintain more than one UPoP identity and collect a UBI in more than one currency (a Sybil attack).

\subsection{Registration}\label{ceremonyprep}
At least 24h before the start of a ceremony, participant $a$ creates a registration transaction for ceremony $i$ containing
\begin{description}
\item [$K_{a,i}^{pub}$] one-time public key
\item [$S_{C,a,j,m}$] (\emph{optional}) a proof that $a$ attended a meetup $m$ at a past ceremony $j$ successfully for the same currency $C$ (reputation).
\item [$C_k$] The local currency's identifier
\end{description}

The registration data is transferred and stored confidentially to mitigate \emph{linkability} accross ceremonies.
\subsection{Assignment}

For each local currency $C_k$, the system then assigns people from the set of registered participants $A_{C,i}$ to meetups $m_{C,i,x}$ at meetup locations from $L_C$ 24h prior to the ceremony.

For participant convenience it would be tempting to let them choose the meetup location themselves. This, however, would open a collusion attack vector because a group of attackers could gain a majority at selected meetups easily. We therefore choose to randomize assignments among $L_C$.

As a mitigation against censorship or sabotage it should not be easy to predict where meetups will take place. Therefore, the set of available locations should be much larger than the number of meetups assigned: $|L_C| \gg |m_{C,i}|$. Moreover, the assignment needs to be kept confidential. Confidentiality is achieved by leveraging SubstraTEE \cite{substraTEE}.

The assignment algorithm will be randomized subject to the following rules: 

\begin{erule}\label{rule:notseeagain}
Minimize the number of participants that have met at their last ceremony already.
\end{erule}

\begin{erule}\label{rule:meetupsize}
Maximize the number of participants per meetup $N_m$ subject to $3 \leq N_m \leq 12\  \forall m$. 
\end{erule}

\begin{erule}\label{rule:noreplimit}
No meetup may be assigned with more than 1/4 of participants without reputation.
\end{erule}
	
\begin{erule}
Randomize meetup locations $l_m$ based on private randomness obtained from a TEE's secure random source.
\end{erule}

The lower limit to meetup size of 3 participants shall provide some safety (see \ref{safetythreat}). 
The upper limit of 12 attendees shall make sure that meetups can be taken out within short time and to allow each participant to remember who she/he has already signed keys with.

Rule \ref{rule:noreplimit} directly impacts the maximum adoption rate: A population of 1M will take at least $\frac{\log(\frac{1M}{12})}{log(\frac{4}{3})} \approx 40$ ceremonies to build up.

\subsubsection{Computational Complexity}
Rules \ref{rule:notseeagain}, \ref{rule:meetupsize} and \ref{rule:noreplimit} form an np-hard optimization problem. Therefore, the assignment has to be performed off-chain.

\subsection{Meetup Procedure}
Shortly before high sun $T_{C,i}$, each attendee $a_n$ shows up at the assigned location $l_m$ and votes on the number of physically present participants $\nu_{m,n}$ using the \encointer mobile phone app. The participant then broadcasts a \emph{claim of attendance} 

\begin{equation}
\zeta_{m,n} = \left(K_{a,i}^{pub}, i, m, \nu_{m,n} \right)
\end{equation}

\begin{erule}
	$\zeta_{m,n}$ has to be broadcast to all meetup participants within $\Delta t_C$ after $T_i$, as attested by the receivers. No latecomers shall be attested.
\end{erule}

\begin{equation}
\Delta t_C = \frac{min\left(dist\left(l_{C,i}, l_{C,k}\right)\right)}{v_{max}} \forall l_{C,k} \in L_C \setminus l_{C,i} 
\end{equation}

Where $v_{max} = 300km/h$ is an \encointer parameter chosen to make it impractical to attend two adjacent ceremonies.

Following the mutual broadcast of claims, participants then attest the physical presence of all counterparties by pairwise signing each others claims. More precisely, participant $a_r$ returns an attestation $\mathcal{A}_r\left(\zeta_{m,n}\right)$, signed with her private key $K^{priv}_{r,i}$ to participant $a_n$ and vice-versa.

\subsection{Witnessing Phase}
During the \emph{witnessing} phase $T_{C,i}\ until\ T_{C,i} + 24h$, each attendee sends his/her collected attestations to the \encointer chain for validation. The system then registers attestations and individual votes $\nu_n$ subject to a set of rules:

\begin{erule}
	Only consider attestations from \emph{other} participants \emph{assigned to the same meetup $m$} and the same ceremony $i$.
\end{erule}

\subsubsection{Storage Complexity}
The \emph{witnessing} registry has a storage complexity of $O(m*n^2)$ where m is the number of meetups and n the number of participants per meetup.

\subsection{Validation}
At the end of the \emph{witnessing} phase, the system validates all meetups subject to the following rules: 

Let $\mathcal{\bar M}_m$ be the set of participants for meetup $m$ with at least one valid attestation from another member in $\mathcal{\bar M}_m$.

Let $\mathcal{\hat M}_m$ be the set of participants for meetup $m$ with reputation (wo can prove recent attendance with $S_{C,a,j,x} |i-j<2$) and at least one valid attestation for meetup $m$ from another member in $\mathcal{\hat M}_m$.

Let $\hat\nu_m$ be the majority vote among $\mathcal{\hat M}_m$.

\begin{erule}\label{rule:votingcorrectly}
	Only consider participants $a_n \in \mathcal{\bar M}_m$ whose vote $\nu_n = \hat\nu_m$
\end{erule}

\begin{erule}
	Only consider participants who collected at least $|\mathcal{\hat M}_m|-2$ attestations.
\end{erule}
\begin{erule}\label{rule:hasattested}
	Only consider participants $a_n \in \mathcal{\bar M}_m$ who have attested $|\mathcal{\hat M}_m|-2 \leq N_n \leq \hat\nu_m$ participants.
\end{erule}

\subsubsection{Computational Complexity}
Validation has a complexity of $O(m*n^2)$ where m is the number of meetups and n the number of participants per meetup. 

\subsection{Reward}
All participants who passed the validation above are issued an amount of $\mathcal{R}_C$ as a universal basic income. $\mathcal{R}_C$ can be defined and adjusted per local currency by means of on-chain governance.  

\subsection{Unique Proof-of-Personhood}
Attending one meetup supplies the individual with a simple proof-of-personhood (PoP). However, it doesn't prove that one individual maintains exactly one PoP over time. The more subsequent ceremonies an individual attends, the more trustworthy is his/her Unique-PoP (UPoP) claim with respect to uniqueness.

\subsection{Threat Model}
The \encointer UPoP protocol needs to defend against two categories of adversaries:
\begin{itemize}
	\item those who try to get more than one reward per ceremony (sybil attack)
	\item those who try to sabotage the \encointer ecosystem even if this comes at a cost.
\end{itemize}
Both categories will collude among themselves to achieve their goal.

\begin{hypothesis}\label{hypothesis:secureifmajorityhonest}
	The \encointer UPoP Protocol is secure if a majority of participants with reputation for each ceremony and each meetup is honest and successfully registers their non-empty attested claims to the blockchain in time. 
\end{hypothesis}

Above hypothesis as not formally proved here. We expect to get probabilistic security with regard to the following attacks:

\subsubsection{Illegit Videoconference}
People may try to meet virtually instead of physically. \emph{Mitigated subject to Hypothesis \ref{hypothesis:secureifmajorityhonest}}
\subsubsection{Surrogates}
An adversary might pay other people to attend ceremonies on behalf of identities controlled by the adversary. The effect is similar to people renting out their identity. \emph{This attack is out of scope as it doesn't affect issuance.}
\subsubsection{Oversigning / Social Engineering}
Attendees might talk others into signing more than one pseudonym per person. Bribery could happen too. \emph{Mitigated by rules \ref{rule:votingcorrectly} and \ref{rule:hasattested} subject to Hypothesis \ref{hypothesis:secureifmajorityhonest}}
\subsubsection{Systematic No-Show}
a meeting might become invalid if too many participants don't show up (deliberately). \emph{Mitigated subject to Hypothesis \ref{hypothesis:secureifmajorityhonest}} 
\subsubsection{Flooding}
An adversary could register large numbers of fake participants who will never show up. This could prevent legit persons to participate.
 \textit{Mitigated by rule \ref{rule:noreplimit}}.
\subsubsection{Threats to Personal Safety}\label{safetythreat}
As ceremony members need to meet in person, all risks involved with human encounters apply. These risks are reduced by randomizing participants and by the minimal group size of 3 persons. Participants are advised to choose public places for ceremonies.
Threats by non-participants who want to hurt the \encointer ecosystem by attacking participants are mitigated if group $s$ keeps their exact meeting point private. 

%%%%%%%%%%%%%%%%%%%%%%%%%%%%%%%%%%%%%%%%%%%%%%%
\section{Monetary Policy}
Unlike Bitcoin, \encointer local currencies don't have a hard-capped supply. The more people are joining the ecosystem, the more money is issued. Every PoP-ceremony participant will receive one reward per attended ceremony as an unconditional basic income (UBI). 

To analyze the macroeconomics of such a basic income scheme let's look at the steady state of a stationary economy with a fixed population and no economic growth. 

If we'd just issue a fixed reward after every ceremony, constant absolute nominal inflation would result. The real value of one (constant) reward would decrease over time and our UBI would become meaningless. If we want our UBI to maintain meaningful value, we can either introduce increasing rewards which may be negatively percieved as hyperinflation. Or we introduce a demurrage fee \cite{gesell}, constantly ''burning'' money. Demurrage has been tested with local currencies \cite{WIR}, \cite{Chiemgauer} and the cryptocurrency Freicoin \cite{freicoin}. \encointer will employ demurrage as a nominally deflationary measure to compensate for the inflation due to ceremonies. This way, a parametrizable percentage of the money supply will be frequently redistributed in egalitarian fashion. The demurrage fee can thus be understood as a tax to the ''decentralized state'' that performs the redistribution of wealth.

\begin{figure}
	\centering
	\def\svgwidth{\columnwidth}
	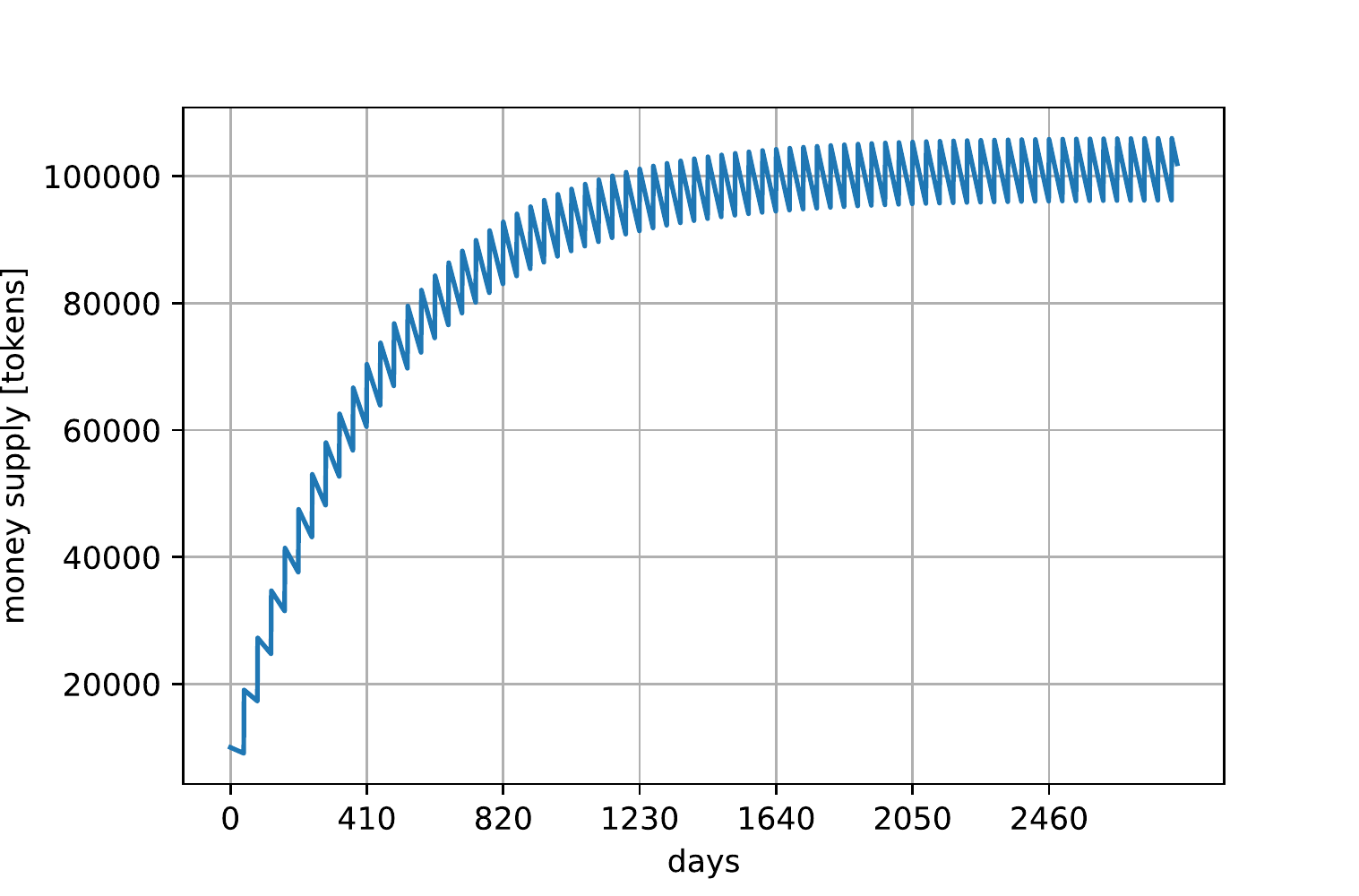
	\caption{Money supply for a stationary economy with a demurrage fee of 7\% per month and a population of 10'000 participants. After an initial phase, the UBI of 1 token per ceremony maintains a constant ratio to the total money supply and therefore maintains a constant real value.}
	\label{fig:ubi-demurrage}
\end{figure}

Fig \ref{fig:ubi-demurrage} shows our example economy reaching an equilibrium state. For an example demurrage of 7\% per month, the money supply saturates at around 100'000 tokens. Such a demurrage causes a deliberate incentive to spend as quickly as possible instead of saving. A high money velocity can be expected. For simplicity, we assume a money velocity of one per ceremony (the sum of money changing hands between two subsequent ceremonies equals the total money supply). With a stable money supply around 100'000 tokens the total of our ceremony rewards (10'000 tokens) will maintain a constant real value of 10\% of all the spending during one period. That ratio will decrease with decreasing rates of demurrage.

Real economies are not stationary and it is not obvious, what a good amount of UBI would be. \encointer therefore allows local currencies to choose their parameters freely by means of on-chain governance. 

%Exponential community growth causes exponential growth of money supply. If community participation should one day saturate, money issuance will be constant and relative inflation rate will therefore decrease over time as shown in figure \ref{fig:inflation}.

%With such a policy in place, no early adopter should expect to get rich by hoarding as adoption drives inflation and demurrage will further disincentivize saving. 

One must expect that people with idle currency will try to avoid demurrage and buy assets better suited to store value, thereby reducing the real value of the \encointer currency. A nation state can enforce a legal tender and force citizens to pay taxes in national currency. An \encointer communitiy on the other hand could promote the use of its currency by social pressure. 

Another way of applying this technology would be to use \encointer tokens as trusted vouchers for cash transfer schemes in fiat currency, run by a state or by NGOs. In this case demurrage should be disabled and replaced by verifiably burning tokens upon exanging them for fiat.

\section{Purchasing-Power Adjusted Transaction Fees}\label{sec:fees}

Bitcoin transactions pay a fee to the miner including the transaction in his block. Because Bitcoin has a hard cap on block size, a market develops for fees and miners choose the best-paying transactions to fill a block. This global fee market restricts adoption in developing countries where many people have very little purchasing power in global perspective. While IOTA \cite{iota} and Nano \cite{nano} have zero transaction fees they must use a small PoW as a measure against spam transactions. Even though small, PoW limits the possibilities to use mobile or IoT devices to send transactions.

Because \encointer currencies are local with an independent valuation, fees can be charged according to purchasing power within each community, as a percentage of the UBI which everybody can get. These fees in local currency are charged by the TEE sidechain validators who maintain the community shard. 

%\encointer targets a balance of power among multiple TEE vendors as it is unlikely that different vendors collude or show the same vulnerabilities. Such a balance of power must be incentivized in order to take place. One possible incentive would be to burn a fraction of tx fees that is proportional to the network share of the validor's TEE vendor. Like this, minority TEEs would earn more fees than majority TEEs.

\section{Architecture}

In this section we describe the system architecture for \encointer as a Polkadot parachain (or parathread) using the SubstraTEE framework. Explaining these technologies is beyond the scope of this paper and the reader shall refer to \cite{polkadot} and \cite{substraTEE}.

\begin{figure}
	\centering
	\def\svgwidth{\columnwidth}
	\includegraphics[width=\columnwidth]{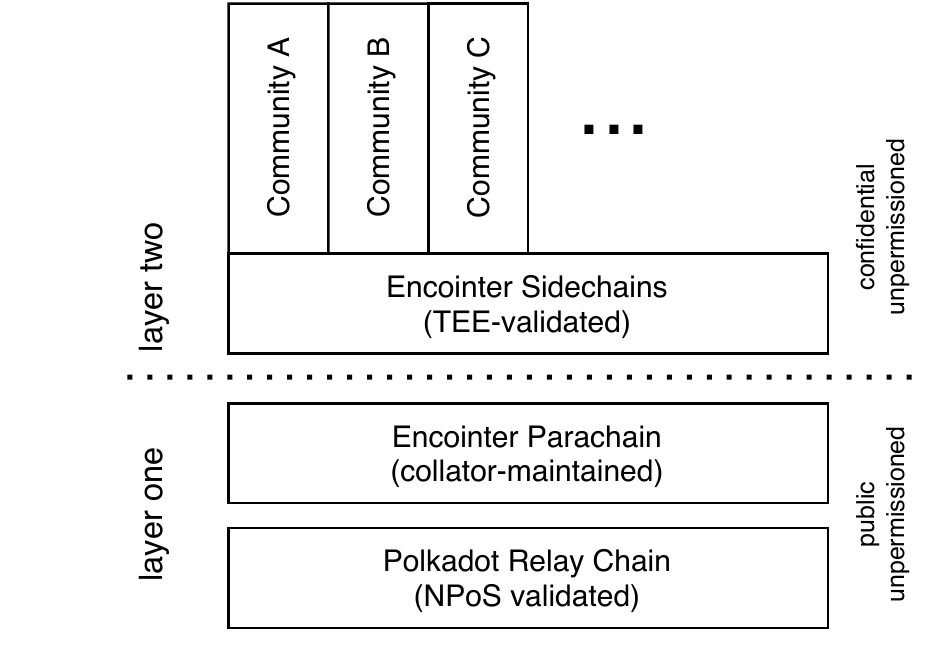}
	\caption{\encointer Stack as a Polkadot parachain using the SubstraTEE framework}
	\label{fig:architecture}
\end{figure}

Figure \ref{fig:architecture} shows the entire stack of the \encointer platform. 

\subsection{Relay Chain}
At the lowest level, we have the Polkadot relay chain as a NPoS-validated root-of-trust. The relay chain's native token is the DOT and the block time is 6s. 

Parachains need to bond DOTs in order to get a slot 

\subsection{Parachain}
The \encointer parachain enjoys the security of the underlying relay chain and serves the following purposes: 

\begin{itemize}
	\item registry for remote-attested TEEs who provide sidechain validation
	\item finality for sidechain blocks
	\item global scheduling of ceremonies
	\item registry of community currencies and their parameters
\end{itemize}

There is no need for the parachain to have its own native token, therefore it will use DOT for fee payments. Sidechain validators will have to pay parachain fees in DOT to have their TEE attested and their blocks finalized. 
Users will not usually interact with the parachain directly but fees for registering a new currency have to be paid in DOT as well. 

Parachain fees are expected to be marginal as they only serve the purpose of spam-prevention. There are no validators that need to be incentivized. While \emph{collators} are needed for availability to produce blocks which are then validated by relay-chain validators, their role is not security critical. \encointer communities will have a strong intrinsic incentive to run collators without much  compensation.

It should be said that there is no guarantee that \encointer can become a parachain as the slots are limited. Auctions where DOT holders can bond their DOTs for parachain candidates will determine slot allocations. Should \encointer not win a slot, it could still run as a \emph{parathread} where DOTs have to be paid for each validated block. The confirmation time for finality will be much longer in that case, but sidechains would not be affected much.

\subsection{Sidechains}

The \encointer sidechains maintain the community currency balance ledger and perform the \encointer protocol for uPoP (See \ref{sec:upop}). Each community has its own shard holding its confidential state.

The sidechain validators are run in TEEs to get both integrity and confidentiality. As TEEs are considered to be trusted, there is no need for a consensus protocol for sidechain blocks. This simplifies the design significantly and allows for sub-second block times due to a low communication overhead.

Sidechain validators can only operate a limited number of shards per machine. Therefore we expect local communities to take care of their own sidechain validator infrastructure.  

Running sidechain validators for any shard is unpermissioned. Every remote-attested TEE will be provisioned with the necessary keys to participate in block-production. All sidechain validators run the same code which is asserted by remote attestation.  

\section{Trusted Execution Environment Security}

TEEs aim to provide the necessary guarantees for secure remote computation. They should provide integrity and confidentiality guarantees when executing software on a computer maintained by an untrusted party. The most recent TEEs rely on software attestation, a process that guarantees the user that she's communicating with a known piece of code running inside a secure container on a genuine trusted hardware by means of a manufacturer signature.

As criticized in \cite{costan16}, manufacturers seem to follow a security by obscurity principle not disclosing design internals necessary for a proper security review. Their \textit{in dubio contra reum} analysis of Intel SGX shows vulnerabilities to cache timing and sidechannel attacks. \textit{Foreshadow} \cite{foreshadow} falsified confidentiality as well as integrity claims for SGX but the attack is mitigated for now. ARM TrustZone on the other hand is only an IP core and design details are left to the manufacturer, equally reluctant to disclose details. 

Since at least the post-Snowden era, one also has to be concerned about manufacturers being forced by their state to introduce deliberate backdoors. Even if open-source TEEs like Keystone \cite{keystone} might soon deliver devices, one would still have to trust the manufacturer not to tamper with the design. 

While all this is disturbing, it should be put in perspective. Information security is a never-ending race. All blockchain solutions are software running by large part on Intel CPUs. While hardware wallets may give us some comfort concerning our funds private keys, there's no guarantee on confidentiality when considering sidechannel attacks. 

The \encointer association will follow developments closely and maintain an up to date list of accepted TEE manufacturers' attestation keys. 

\section{\encointer Association}

The \encointer association is a not-for-profit association under Swiss law. Its purpose is to govern the \encointer ecosystem during its initial phase. It fulfills the following tasks
\begin{itemize}
	\item community bootstrapping
	\item protocol updates
	\item maintain list of accepted TEE attestation service keys
\end{itemize}

All changes are subject to a referendum by the community.

\section{Known Limitations}
\subsection{Scalability}
The proposed \encointer protocol assumes that the entire state for a local community can fit into secure memory within a TEE. This limits the number of accounts that can be registered per community. 

\section{Conclusion}
A novel cryptocurrency system has been introduced in conceptual detail. Main contributions are 
\begin{itemize}
	\item A novel approach to monetary policy supporting equal opportunity globally with a  universal basic income (UBI) in local community currencies.
	\item A novel protocol for trustless pseudonym key signing parties for proof-of-personhood (PoP)
	\item Private transactions with purchasing-power adjusted fees.
\end{itemize}
%%%%%%%%%%%%%%%%%%%%%%%%%%%%%%%%%%%%%%%%%%%%%%%%%%%%%%%
%%%%%%%%%%%%%%%%%%%%%%%%%%%%%%%%%%%%%%%%%%%%%%%%%%%%%%

\end{document}